# Elsevier instructions for the preparation of a 2-column format camera-ready paper in LaTeX

P. de Groot[a]* and X.-Y. Wang[b]

[a]Mathematics and Computer Science Division, Elsevier Science Publishers B.V.,
P.O. Box 103, 1000 AC Amsterdam, The Netherlands

[b]Economics Department, University of Winchester,
2 Finch Road, Winchester, Hampshire P3L T19, United Kingdom

These pages provide you with an example of the layout and style for 100% reproduction which we wish you to adopt during the preparation of your paper. This is the output from the LaTeX document class you requested.

## 1. FORMAT

Text should be produced within the dimensions shown on these pages: each column 7.5 cm wide with 1 cm middle margin, total width of 16 cm and a maximum length of 20.2 cm on first pages and 21 cm on second and following pages. The LaTeX document class uses the maximal stipulated length apart from the following two exceptions (i) LaTeX does not begin a new section directly at the bottom of a page, but transfers the heading to the top of the next page; (ii) LaTeX never (well, hardly ever) exceeds the length of the text area in order to complete a section of text or a paragraph. Here are some references: [1,2].

### 1.1. Spacing

We normally recommend the use of 1.0 (single) line spacing. However, when typing complicated mathematical text LaTeX automatically increases the space between text lines in order to prevent sub- and superscript fonts overlapping one another and making your printed matter illegible.

### 1.2. Fonts

These instructions have been produced using a 10 point Computer Modern Roman. Other recommended fonts are 10 point Times Roman, New Century Schoolbook, Bookman Light and Palatino.

## 2. PRINTOUT

The most suitable printer is a laser printer. A dot matrix printer should only be used if it possesses an 18 or 24 pin printhead ("letter-quality").

The printout submitted should be an original; a photocopy is not acceptable. Please make use of good quality plain white A4 (or US Letter) paper size. *The dimensions shown here should be strictly adhered to: do not make changes to these dimensions, which are determined by the document class.* The document class leaves at least 3 cm at the top of the page before the head, which contains the page number.

Printers sometimes produce text which contains light and dark streaks, or has considerable lighting variation either between left-hand and right-hand margins or between text heads and bottoms. To achieve optimal reproduction quality, the contrast of text lettering must be uniform, sharp and dark over the whole page and throughout the article.

If corrections are made to the text, print completely new replacement pages. The contrast on these pages should be consistent with the rest of the paper as should text dimensions and font sizes.

---

*Footnotes should appear on the first page only to indicate your present address (if different from your normal address), research grant, sponsoring agency, etc. These are obtained with the \thanks command.

## 3. TABLES AND ILLUSTRATIONS

Tables should be made with LaTeX; illustrations should be originals or sharp prints. They should be arranged throughout the text and preferably be included *on the same page as they are first discussed*. They should have a self-contained caption and be positioned in flush-left alignment with the text margin within the column. If they do not fit into one column they may be placed across both columns (using `\begin{table*}` or `\begin{figure*}` so that they appear at the top of a page).

### 3.1. Tables

Tables should be presented in the form shown in Table 1. Their layout should be consistent throughout.

Horizontal lines should be placed above and below table headings, above the subheadings and at the end of the table above any notes. Vertical lines should be avoided.

If a table is too long to fit onto one page, the table number and headings should be repeated above the continuation of the table. For this you have to reset the table counter with `\addtocounter{table}{-1}`. Alternatively, the table can be turned by 90° ('landscape mode') and spread over two consecutive pages (first an even-numbered, then an odd-numbered one) created by means of `\begin{table}[h]` without a caption. To do this, you prepare the table as a separate LaTeX document and attach the tables to the empty pages with a few spots of suitable glue.

### 3.2. Useful table packages

Modern LaTeX comes with several packages for tables that provide additional functionality. Below we mention a few. See the documentation of the individual packages for more details. The packages can be found in LaTeX's `tools` directory.

**array** Various extensions to LaTeX's `array` and `tabular` environments.

**longtable** Automatically break tables over several pages. Put the table in the `longtable` environment instead of the `table` environment.

Table 2: The next-to-leading order (NLO) results *without* the pion field.

| $\Lambda$ (MeV) | 140 | 150 | 175 | 200 | 225 | 250 | Exp. | $v_{18}$ [?] |
|---|---|---|---|---|---|---|---|---|
| $r_d$ (fm) | 1.973 | 1.972 | 1.974 | 1.978 | 1.983 | 1.987 | 1.966(7) | 1.967 |
| $Q_d$ (fm$^2$) | 0.259 | 0.268 | 0.287 | 0.302 | 0.312 | 0.319 | 0.286 | 0.270 |
| $P_D$ (%) | 2.32 | 2.83 | 4.34 | 6.14 | 8.09 | 9.90 | — | 5.76 |
| $\mu_d$ | 0.867 | 0.864 | 0.855 | 0.845 | 0.834 | 0.823 | 0.8574 | 0.847 |
| $\mathcal{M}_{M1}$ (fm) | 3.995 | 3.989 | 3.973 | 3.955 | 3.936 | 3.918 | — | 3.979 |
| $\mathcal{M}_{GT}$ (fm) | 4.887 | 4.881 | 4.864 | 4.846 | 4.827 | 4.810 | — | 4.859 |
| $\delta_{1B}^{VP}$ (%) | −0.45 | −0.45 | −0.45 | −0.45 | −0.45 | −0.44 | — | −0.45 |
| $\delta_{1B}^{C2:C}$ (%) | 0.03 | 0.03 | 0.03 | 0.03 | 0.03 | 0.03 | — | 0.03 |
| $\delta_{1B}^{C2:N}$ (%) | −0.19 | −0.19 | −0.18 | −0.15 | −0.12 | −0.10 | — | −0.21 |

The experimental values are given in ref. [4].

Table 1
The next-to-leading order (NLO) results *without* the pion field.

| $\Lambda$ (MeV) | 140 | 150 | 175 | 200 |
|---|---|---|---|---|
| $r_d$ (fm) | 1.973 | 1.972 | 1.974 | 1.978 |
| $Q_d$ (fm$^2$) | 0.259 | 0.268 | 0.287 | 0.302 |
| $P_D$ (%) | 2.32 | 2.83 | 4.34 | 6.14 |
| $\mu_d$ | 0.867 | 0.864 | 0.855 | 0.845 |
| $\mathcal{M}_{\mathrm{M1}}$ (fm) | 3.995 | 3.989 | 3.973 | 3.955 |
| $\mathcal{M}_{\mathrm{GT}}$ (fm) | 4.887 | 4.881 | 4.864 | 4.846 |
| $\delta_{\mathrm{1B}}^{\mathrm{VP}}$ (%) | $-0.45$ | $-0.45$ | $-0.45$ | $-0.45$ |
| $\delta_{\mathrm{1B}}^{\mathrm{C2:C}}$ (%) | 0.03 | 0.03 | 0.03 | 0.03 |
| $\delta_{\mathrm{1B}}^{\mathrm{C2:N}}$ (%) | $-0.19$ | $-0.19$ | $-0.18$ | $-0.15$ |

The experimental values are given in ref. [4].

`dcolumn` Define your own type of column. Among others, this is one way to obtain alignment on the decimal point.

`tabularx` Smart column width calculation within a specified table width.

`rotating` Print a page with a wide table or figure in landscape orientation using the `sidewaystable` or `sidewaysfigure` environments, and many other rotating tricks. Use the package with the `figuresright` option to make all tables and figures rotate in clockwise. Use the starred form of the `sideways` environments to obtain full-width tables or figures in a two-column article.

### 3.3. Line drawings

Line drawings should be drawn in India ink on tracing paper with the aid of a stencil or should be glossy prints of the same; computer prepared drawings are also acceptable. They should be attached to your manuscript page, correctly aligned, using suitable glue and *not transparent tape*. When placing a figure at the top of a page, the top of the figure should be at the same level as the bottom of the first text line.

All notations and lettering should be no less than 2 mm high. The use of heavy black, bold lettering should be avoided as this will look unpleasantly dark when printed.

### 3.4. PostScript figures

Instead of providing separate drawings or prints of the figures you may also use PostScript files which are included into your LaTeX file and printed together with the text. Use one of the packages from LaTeX's `graphics` directory: `graphics`, `graphicx` or `epsfig`, with the `\usepackage` command, and then use the appropriate commands (`\includegraphics` or `\epsfig`) to include your PostScript file.

The simplest command is:
`\includegraphics{file}`, which inserts the PostScript file `file` at its own size. The starred version of this command:
`\includegraphics*{file}`, does the same, but clips the figure to its bounding box.

With the `graphicx` package one may specify a series of options as a key–value list, e.g.:
`\includegraphics[width=15pc]{file}`
`\includegraphics[height=5pc]{file}`
`\includegraphics[scale=0.6]{file}`
`\includegraphics[angle=90,width=20pc]{file}`

See the file `grfguide`, section "Including Graphics Files", of the `graphics` distribution for all options and a detailed description.

The `epsfig` package mimicks the commands familiar from the package with the same name in LaTeX2.09. A PostScript file `file` is included with the command `\psfig{file=file}`.

Grey-scale and colour photographs cannot be included in this way, since reproduction from the printed CRC article would give insufficient typographical quality. See the following subsections.

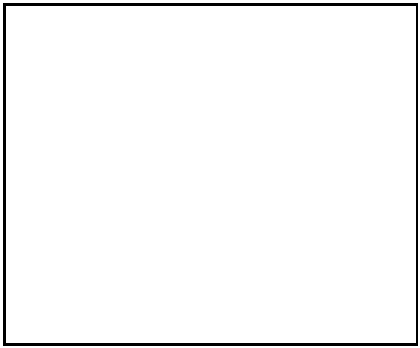

Figure 1. Good sharp prints should be used and not (distorted) photocopies.

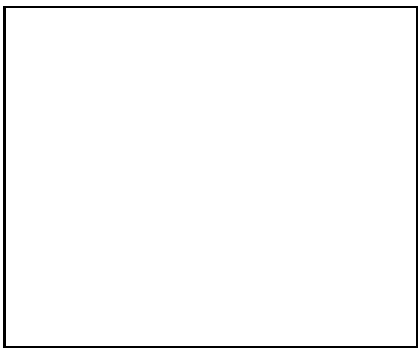

Figure 2. Remember to keep details clear and large enough.

### 3.5. Black and white photographs

Photographs must always be sharp originals (*not screened versions*) and rich in contrast. They will undergo the same reduction as the text and should be pasted on your page in the same way as line drawings.

### 3.6. Colour photographs

Sharp originals (*not transparencies or slides*) should be submitted close to the size expected in publication. Charges for the processing and printing of colour will be passed on to the author(s) of the paper. As costs involved are per page, care should be taken in the selection of size and shape so that two or more illustrations may be fitted together on one page. Please contact the Technical Editor in the Camera-Ready Publications Department at Elsevier for a price quotation and layout instructions before producing your paper in its final form.

## 4. EQUATIONS

Equations should be flush-left with the text margin; LaTeX ensures that the equation is preceded and followed by one line of white space. LaTeX provides the package `fleqn` to get the flush-left effect.

$$H_{\alpha\beta}(\omega) = E_\alpha^{(0)}(\omega)\delta_{\alpha\beta} + \langle\alpha|W_\pi|\beta\rangle \qquad (1)$$

You need not put in equation numbers, since this is taken care of automatically. The equation numbers are always consecutive and are printed in parentheses flush with the right-hand margin of the text and level with the last line of the equation. For multi-line equations, use the `eqnarray` environment. For complex mathematics, use the $\mathcal{AMS}$-LaTeX package.

References should be collected at the end of your paper. Do not begin them on a new page unless this is absolutely necessary. They should be prepared according to the sequential numeric system making sure that all material mentioned is generally available to the reader. Use `\cite` to refer to the entries in the bibliography so that your accumulated list corresponds to the citations made in the text body.

Above we have listed some references according to the sequential numeric system [1–4].

# Partonic structure of the virtual photon [*]

J. Chýla and M. Taševský [a]

[a]Institute of Physics of the Academy of Sciences,
Na Slovance 2, Prague 8, Czech Republic

Interactions of virtual photons are analyzed in terms of their parton distribution functions. It is shown that the concept of parton distribution functions is phenomenologically very useful even for highly virtual photons involved in hard collisions. The role of longitudinal photons for proper interpretation of the data on jet cross–section in the region of moderate photon virtualities accessible at HERA is explored.

## 1. Introduction

In quantum field theory it is difficult to distinguish effects of the "structure" from those of "interactions". Within the Standard Model it makes good sense to distinguish *fundamental particles*, which correspond to fields in its lagrangian $\mathcal{L}_{\rm SM}$ (leptons, quarks, gauge and Higgs bosons) from *composite particles*, like atoms and hadrons, which appear in the mass spectrum but have no corresponding fields in $\mathcal{L}_{\rm SM}$. For the latter the use of parton distribution functions (PDF) to describe their "structure" appears natural, but the concept of PDF turns out to be phenomenologically useful also for some fundamental particles, in particular the photon.

## 2. PDF of the real photon

The factorization scale dependence of PDF of the real photon is determined by the system of coupled inhomogeneous evolution equations

$$\frac{{\rm d}\Sigma(M^2)}{{\rm d}\ln M^2} = k_q + P_{qq} \otimes \Sigma + P_{qG} \otimes G, \quad (1)$$

$$\frac{{\rm d}q_{NS}(M^2)}{{\rm d}\ln M^2} = \sigma_{\rm NS}k_q + P_{\rm NS} \otimes q_{\rm NS}, \quad (2)$$

$$\frac{{\rm d}G(M^2)}{{\rm d}\ln M^2} = k_G + P_{Gq} \otimes \Sigma + P_{GG} \otimes G \quad (3)$$

for singlet, nonsinglet and gluon distribution functions. The spliting functions $k_q, k_G$ and $P_{ij}$ admit expansions in powers of $\alpha_s$, which start at the order $\alpha\alpha_s$, except for $k_q$ which starts as $(\alpha/2\pi)3e_q^2(x^2 + (1-x)^2) = O(\alpha)$. The general solution of these equations can be written as the sum of a particular solution of the full inhomogeneous equations and the general solution of the corresponding homogeneous ones, called *hadronic* (or VDM) part. A subset of solutions of the inhomogeneous evolution equations resulting from the resummation of diagrams in Fig. 1 defines the so called *pointlike* (PL) parts. This resummation softens the $x$−dependence of $q^{\rm PL}(x, M^2)$ with respect to the term corresponding to the simple $\gamma \to q\overline{q}$ splitting. A general solution of eqs. (1-3) can thus be written as ($D = q, \overline{q}, G$)

$$D(x, M^2) = D^{\rm PL}(x, M^2) + D^{\rm VDM}(x, M^2). \quad (4)$$

The important point to keep in mind is the fact that there is an infinite number of pointlike solutions $q^{\rm PL}(x, M^2)$, $G^{\rm PL}(x, M^2)$, differing by the initial scale $M_0$ at which they vanish. Consequently, the separation of quark and gluon distribution functions into their pointlike and hadronic parts is ambiguous and therefore these concepts have separately no physical meaning [2].

Practical aspects of the ambiguity in separating PDF into their VDM and PL parts can be nicely illustrated on the properties of SaS1D and SaS2D parametrizations [1,2], which provide separate parametrizations of the VDM and PL parts of both quark and gluon distributions and which differ by the choice of initial $M_0$: $M_0 = 0.6$ GeV for SaS1D, $M_0 = 2$ GeV for SaS2D.

---

[2]For brevity the terms "pointlike quarks" and "pointlike gluons" will hence be employed to denote pointlike parts of quark and gluon distribution functions of the photon.

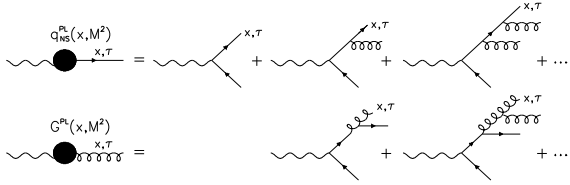

Figure 1. *Diagrams defining the pointlike parts of quark and gluon distribution functions in LL approximation. The resummation involves integration over parton virtualities $M_0^2 \leq \tau \leq M^2$.*

In [3] we have compared distribution functions $xu(x, M^2)$, $xc(x, M^2)$ and $xG(x, M^2)$ as given by SaS1D and SaS2D parametrizations as functions of $x$ and $M^2$ and identified the effects of the resummation of multiple parton emission on the simple splitting term $q^{\text{split}}(x, M_0^2, M^2) = (\alpha/2\pi)3e_q^2(x^2 + (1-x)^2) \ln(M^2/M_0^2)$. We also investigated the scale dependence of VDM and PL parts of these distribution functions and compared individual contributions (quark vs. gluon, PL. vs VDM) to LO predictions for two physical quantities: $F_2^\gamma(x, Q^2) = \sum_{i=1}^{n_f} 2xe_i^2 q_i(x, Q^2)$ and the so called effective PDF $D_{\text{eff}}(x, M^2) = \sum_{i=1}^{n_f} \left( q_i(x, M^2) + \bar{q}_i(x, M^2) \right) + \frac{9}{4} G(x, M^2)$, relevant for approximate calculations of jet production. These studies (see [3] for figures illustrating these claims) show that

- There is a huge difference between the role of VDM components in SaS1D and SaS2D parametrizations: for SaS2D the VDM parts of $xu(x, M)$ and $xG(x, M)$ dominate up to $x \doteq 0.75$, whereas for SaS1D the PL one dominates already above $x \doteq 0.1$!

- Factorization scale dependence of VDM and PL parts differ substantially. VDM components exhibit the pattern of scaling violations typical for hadrons, whereas the pointlike ones rise, for quarks as well as gluons, with $M$ for all $x$.

- As the factorization scale $M$ increases the VDM parts of both quark and gluon distribution functions decrease relative to the pointlike ones, except for very small $x$.

- Despite huge differences between SaS1D and SaS2D parametrizations in the decomposition of quark and gluon distributions into their VDM and PL parts, their predictions for $F_2^\gamma$ and $D_{\text{eff}}$ are quite close.

- The most prominent effect of multiple parton emission on physical quantities appears to be the contribution of pointlike gluons to jet cross–sections.

## 3. PDF of the virtual photon

For the virtual photon the initial state singularity resulting from the splitting $\gamma^* \to q\bar{q}$ is shielded off by the nonzero photon virtuality $P^2$ and therefore in principle the concept of PDF does not have to be introduced. Nevertheless, even in such circumstances PDF turn out to be very useful phenomenologically because their PL parts include the resummation of parts of higher order QCD corrections, and the VDM ones, though decreasing rapidly with increasing $P^2$, are still dominant at very small $x_\gamma$. Both of these aspects define the "nontrivial" structure of the virtual photon in the sense that they are not included in existing NLO unsubtracted direct photon calculations.

In QCD the nonperturbative effects connected with the confinement are expected to determine the long–range structure of the photon and hence also the transition between the virtual and real photon. As for the real photon, we recall basic features of SaS parametrizations of PDF of the virtual photon and refer the reader to [3,4] for detailed justification.

- Both VDM and PL parts drop with increasing $P^2$, but VDM parts drop much faster.

- With increasing $P^2$ the importance of VDM parts of both quark and gluon distribution functions drops rapidly. For $M^2 \geq 25$ GeV$^2$ the VDM parts of both SaS1D and SaS2D parametrizations become practically negligible already at $P^2 \approx 3$ GeV$^2$, except in the region of very small $x$. Hence, also the ambiguity in the separation (4) is practically irrelevant in this region.

- The general pattern of scaling violations remains the same as for the real photon, except for a subtle difference (see Fig. 3 of [3]) reflecting the fact that SaS parametrizations of PDF of the virtual photon do not satisfy the same evolution equations as PDF of the real one.

- Pointlike quarks dominate $D_{\text{eff}}(x, P^2, M^2)$ at large $x$, while for $x \lesssim 0.5$, most of the pointlike contribution comes from pointlike gluons.

- For $x \gtrsim 0.6$ the full results are below those given by the splitting term with $M_0^2 = P^2$ and one therefore expects the sum DIR+RES to be smaller than the results of unsubtracted direct calculations.

Jet production in $ep$ collisions in the region of photon virtualities $P^2 \gtrsim 1$ offers thus a promising opportunity for the identification of nontrivial aspects of PDF of virtual photons at both small (but not very small) and large values of $x$.

## 4. Should we care about $\gamma_L^*$?

Most of the existing phenomenological analyses of the properties and interactions of virtual photons as well as all available parametrizations of their PDF concern transverse photons only. Neglecting longitudinal photons is a good approximation for $y \to 1$, where the flux $f_L^\gamma(y, P^2) \to 0$, as well as for very small virtualities $P^2$, where PDF of $\gamma_L^*$ vanish by gauge invariance. But how small is "very small" in fact? For instance, should we take into account the contribution of $\gamma_L^*$ to jet cross–sections in the region $E_T^{\text{jet}} \gtrsim 5$ GeV, $P^2 \gtrsim 1$ GeV$^2$, where most of the data on virtual photons extracted from $ep$ collisions at HERA come from? Simple QED based estimates of their effects suggest that in the mentioned kinematical region $\gamma_L^*$ must be taken into account in the resolved photon contribution but may be safely neglected in the direct one. This difference comes from the fact that at small $P^2$ the contributions of $\gamma_L^*$ to physical cross–sections behave as $P^2/\hat{s}$ (i.e. vanish for fixed $P^2$ when $\hat{s} \to \infty$) in the direct channel, but as $P^2/\mu^2$ (with $\mu$ a fixed parameter) in the resolved part. In simple QED based calculations

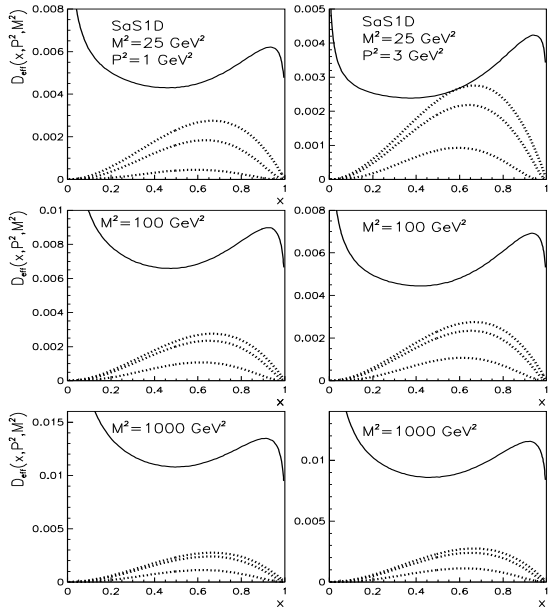

Figure 2. $D_{\text{eff}}(x, P^2, M^2)$ calculated from SaS1D parametrizations for $\gamma_T^*$ (solid curves) compared to results for $\gamma_L^*$ displayed by dotted curves and corresponding from above to $m^2 = 0, 0.1, 1$ GeV$^2$.

[4] $\mu$ is given by quark masses, while in realistic QCD we expect it to be given by some nonperturbative parameter of the order of 1 GeV. To illustrate the importance of $\gamma_L^*$, LO QCD expressions for $D_{\text{eff}}(x, P^2, M^2)$ evaluated with SaS1D parametrizations for $\gamma_T^*$ are compared in Fig. 2 with the formula (44) of [4] for $\gamma_L^*$, treating $m$ in the latter as a free parameter. The dotted curves in this figure correspond (from below) to $m^2 = 1$ GeV$^2$, $m^2 = 0.1$ GeV$^2$ and to the asymptotic expression $q^{\gamma_L^*}(x) = (\alpha/2\pi)12e_q^2x(1-x)$, obtained in the limit $m \to 0$. As expected the importance of $\gamma_L^*$ depends sensitively on $m$. Moreover, its contributions relative to those of $\gamma_T^*$ peak at about $x \approx 0.65$, and drop with increasing $M^2$ (for $P^2$ and $m^2$ fixed), and with increasing $m^2$ (for $P^2$ and $M^2$ fixed).

In Fig. 2 we compared the contributions of $\gamma_T^*$ and $\gamma_L^*$, despite the fact that their respective fluxes differ, but it is trivial to modify the above considerations by taking the respective fluxes properly into account.

## 5. PDF in NLO QCD calculations

The data on dijet production in ep collisions in the region of photon virtualities $1.5 \lesssim P^2 \lesssim 25$ GeV$^2$ and for jet transverse energies $E_T^{\text{jet}} \geq 5$ GeV [5] offer a suitable circumstances for comparison with NLO parton level calculations. In principle this data may be analyzed within the NLO parton level Monte–Carlo programs that do not introduce the concept of PDF of virtual photons (like DISENT, MEPJET or DISASTER++). Nevertheless, so long as $P^2 \ll M^2 \approx E_T^2$, the pointlike parts of PDF incorporate numerically important effects of a part of higher order corrections, namely those coming from collinear emission of partons in Fig. 1. To illustrate this point we shall now discuss the dijet cross–sections calculated by means of JETVIP [6], currently the only NLO parton level Monte–Carlo program that includes both the direct and resolved photon contributions. [3]

Once the concept of virtual photon structure is introduced, part of the direct photon contribution, namely the splitting term (which for the virtual photon is nonsingular), is subtracted from it and included in PDF appearing in the resolved photon contribution. To avoid confusion we shall use the term "direct unsubtracted" (DIR$_{uns}$) to denote NLO direct photon contributions before this subtraction and reserve the term "direct" (DIR) for the results after it. In this terminology the complete calculations is then given by the sum of direct and resolved parts and denoted DIR+RES.

For complete $O(\alpha_s^2)$ calculations only the LO resolved photon contribution must be added to the $O(\alpha \alpha_s^2)$ direct one. However, JETVIP includes also NLO resolved ones. This might seem inconsistent as the corresponding complete $O(\alpha \alpha_s^3)$ direct photon terms are not yet available and cannot thus be included. Nevertheless, this procedure makes sense precisely because of a clear physical meaning of PDF of the virtual photon! The main argument for adding $O(\alpha \alpha_s^3)$ resolved photon terms to $O(\alpha \alpha_s^2)$ direct and $O(\alpha_s^3)$ resolved photon contributions is based on specific way factorization mechanism works for processes involving initial photons. This point is crucial but subtler and we therefore merely summarize the conclusions and refer the reader to [3,7], for details. In fact one can look at $O(\alpha_s^3)$ resolved photon terms as results of approximate evaluation of the so far uncalculated $O(\alpha \alpha_s^3)$ direct photon diagrams in the collinear kinematics. There are of course $O(\alpha \alpha_s^3)$ direct photon contributions that cannot be obtained in this way, but we are convinced that it makes sense to build phenomenology on this framework.

---

[3]In this subsection the various terms considered are characterized by the powers of $\alpha$ and $\alpha_s$ appearing in hard scattering cross–sections. Writing $O(\alpha^j \alpha_s^k)$ will thus mean terms proportional to $\alpha^j \alpha_s^k$, *not* terms *up to* this order! For approximations taking into account the first two or three powers of $\alpha_s$, in either direct or resolved channel, the denomination NLO, NNLO are used.

For the $O(\alpha_s^2)$ resolved terms the so far unknown $O(\alpha \alpha_s^3)$ direct photon contributions provide the first chance to generate pointlike gluons inside the photon. To get the gluon convoluted with $O(\alpha_s^3)$ resolved photon contributions would require evaluating diagrams of even higher order $O(\alpha \alpha_s^4)$! In other words, although the pointlike parts of quark and gluon distribution functions of the virtual photon are in principle included in higher order perturbative corrections and can therefore be considered as expressions of "interactions" rather than "structure", their uniqueness and phenomenological usefulness definitely warrant their introduction as well as their names.

To make our conclusions potentially relevant for ongoing analyses of HERA data we have chosen the following kinematical region (jets with highest and second highest $E_T$ are labelled "1" and "2" and all quantities are in $\gamma^* p$ CMS)

$$E_T^{(1)} \geq 7 \text{ GeV}, E_T^{(2)} \geq 5 \text{ GeV}$$

$$-2.5 \leq \eta^{(i)} \leq 0, \ i = 1, 2$$

and repeatad the analysis in four windows of $P^2$: (1.4, 2.4); (2.4, 4.4), (4.4, 10), (10, 25) GeV$^2$. The cuts were chosen in such a way that throughout the selected region $P^2 \ll E_T^2$, thereby ensuring that the virtual photon lives long enough for its "structure" to develop before the hard scattering takes place. We have chosen the asymmetric cut scenario $E_T^{(1)} \geq E_T^c + \Delta$, $E_T^{(2)} \geq E_T^c$, which avoids the problems coming from the region where $E_T^{(1)} \approx E_T^{(2)}$. To determine the value of $\Delta$ optimally, we evaluated the integral $\sigma(\Delta)$ over the selected region in $E_T^{(1)} - E_T^{(2)}$ plane as a function of $\Delta$ and on the basis thereof took $\Delta = 2$ GeV for all $P^2$.

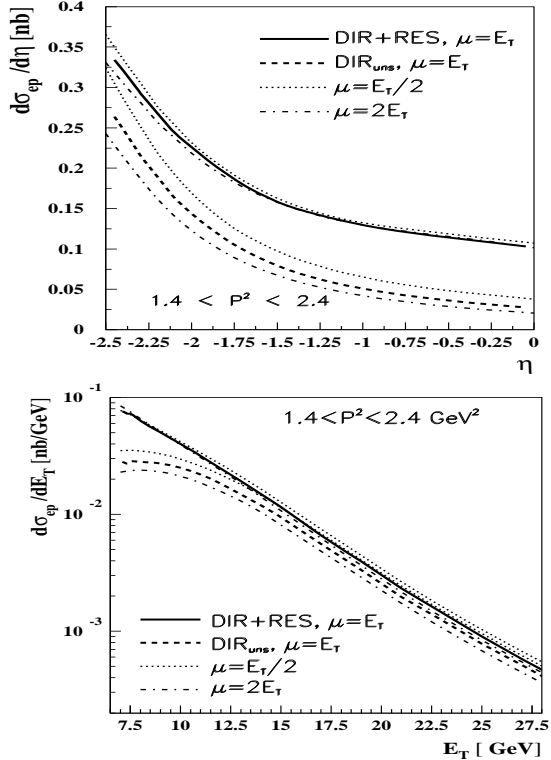

Figure 3. *Scale dependence of* $d\sigma/d\eta$ *and* $d\sigma/dE_T$ *at the NLO with* $R_{sep} = 2R$.

In our analysis jets are defined by means of the cone algorithm. At NLO parton level all jet algorithms are essentially equivalent to the cone one, supplemented with the parameter $R_{sep}$, introduced in order to bridge the gap between the application of the cone algorithm to NLO parton level calculations and to hadronic systems, where one encounters ambiguities concerning the seed selection and jet merging. The question which value of $R_{sep}$ to choose for the comparison of

NLO parton level calculations with the results of the cone algorithm applied at the hadron level is nontrivial and we shall therefore present JETVIP results for both extreme choices $R_{sep} = R$ and $R_{sep} = 2R$. To define momenta of jets JETVIP uses the standard $E_T$–weighting recombination procedure, which leads to massless jets. To assess the reliability of our conclusions we have investigated the following uncertainties:

**Choice of PDF:** We have taken CTEQ4M and SAS1D sets of PDF of the proton and photon respectively as our canonical choice. Both of these sets treat quarks, including $c$ and $b$ ones, as massless above their respective mass thresholds, as required by JETVIP, which uses LO and NLO matrix elements of massless partons. We set $N_f = 4$ in calculations discussed below. PDF of the proton are well determined from global analyses of CTEQ and MRS groups and we have therefore estimated the residual uncertainty related to the choice of PDF of the proton by comparing the CTEQ4M results to those obtained with MRS(2R) set. The differences are tiny, between 3% at $\eta = -2.5$ and 1.5% at $\eta = 0$.

**Factorization scale dependence:** In principle proton and (in resolved channel) photon are associated with different factorization scales $M_p$ and $M_\gamma$, but we followed the standard practice and set $M_p = M_\gamma = M = \kappa(E_T^{(1)} + E_T^{(2)})/2$. The factorization scale dependence was estimated by performing the calculations for $\kappa = 1/2, 1, 2$.

**Renormalization scale dependence:** The dependence of perturbative calculations on the renormalization scale $\mu$ is in principle a separate ambiguity, unrelated to that of $M$, but we followed the common practice and set $\mu = M$.

**Hadronization corrections:** Adopting the definition used by experimentalists [8], we have found that they depended sensitively and in a correlated manner on transverse energies and pseudorapidities of jets. For $E_T^c = 5$ GeV they start to rise steeply below $\eta \doteq -2.5$ and we have therefore required both jets to lie in the region $-2.5 \leq \eta^{(i)} \leq 0$, where hadronization corrections are flat in $\eta$ and do not exceed $10-20\%$. We have run JETVIP in two different modes:

**DIR$_{uns}$:** NLO unsubtracted direct photon cal-culations are performed without introducing the concept of virtual photon structure.

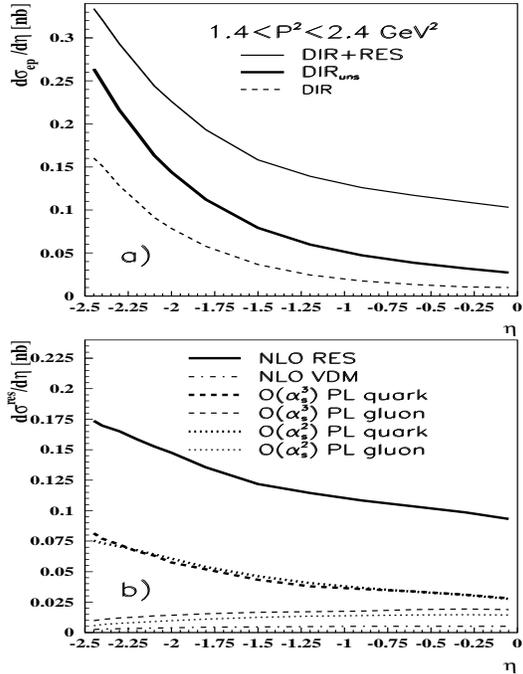

Figure 4. *DIR+RES, DIR$_{uns}$ and DIR results for* d$\sigma$/d$\eta$ *(a) and composition of* d$\sigma$/d/$\eta$ *(b).*

**DIR+RES:** employs the concept of PDF of the virtual photon and gives the jet cross–sections as sums of subtracted direct (DIR) and resolved photon (RES) contributions.

Let us first discuss the results for the window $1.4 \leq P^2 \leq 2.4$ GeV$^2$. In Fig. 3 d$\sigma$/d$\eta$ and d$\sigma$/d$E_T$ distributions of trigger jets [4] and obtained within the DIR+RES approach are compared to those of DIR$_{uns}$ one for $R_{sep} = 2$. The difference between the results of these two approaches is significant in the whole range of $\eta$, but particularly large close to $\eta = 0$, where the DIR+RES results exceed the DIR$_{uns}$ ones by a factor of more than 3! In d$\sigma$/d$E_T$ distributions this difference comes predominantly from the region of $E_T$ close to $E_T^c + \Delta = 7$ GeV. Fig. 3 also shows that the scale dependence is nonnegligible in both approaches, but does not invalidate the main conclusion drawn from this comparison. The dependence of the above results on $R_{sep}$ is almost imperceptible for DIR$_{uns}$ calculations and below 10% for the DIR+RES ones. To track down the origins of the observed large differences between DIR+RES and DIR$_{uns}$ results, we did two exercises. In Fig. 4a the DIR+RES and DIR$_{uns}$ results are compared to the subtracted direct (DIR) ones. The difference between the DIR+RES and DIR curves, defining the resolved photon contribution d$\sigma^{res}$/d$\eta$, is then split into the contributions of:

- the VDM part of photonic PDF convoluted with complete NLO (i.e. up to the order $O(\alpha_s^3)$) parton level cross–sections (denoted NLO VDM),

- the pointlike quark and gluon distribution functions convoluted with $O(\alpha_s^2)$ and $O(\alpha_s^3)$ parton level cross–sections

and plotted in Fig. 4b. We conclude that:

- The contribution of the VDM part of photonic PDF is very small and perceptible only close to $\eta = 0$. Integrally it amounts to about 3%.

- The inclusion of $O(\alpha_s^3)$ resolved photon contributions is numerically important in the whole range $-2.5 \leq \eta \leq 0$. Interestingly, for quarks as well as gluons, the $O(\alpha_s^3)$ results come out quite close to the $O(\alpha_s^2)$ ones.

- At both $O(\alpha_s^2)$ and $O(\alpha_s^3)$ orders pointlike quarks dominate d$\sigma^{res}$/d$\eta$ at large negative $\eta$, whereas as $\eta \to 0$ the fraction of d$\sigma^{res}$/d$\eta$ coming from pointlike gluons increases towards about 40% at $\eta = 0$.

---

[4]In our dijet sample this distribution is equal to the sum of distributions of the first and second jet.

We reiterate that pointlike gluons carry nontrivial information already in convolutions with $O(\alpha_s^2)$ partonic cross–sections because in unsubtracted direct calculations such contributions would appear first at the order $\alpha\alpha_s^3$. In JETVIP the nontrivial aspects of taking into account the resolved

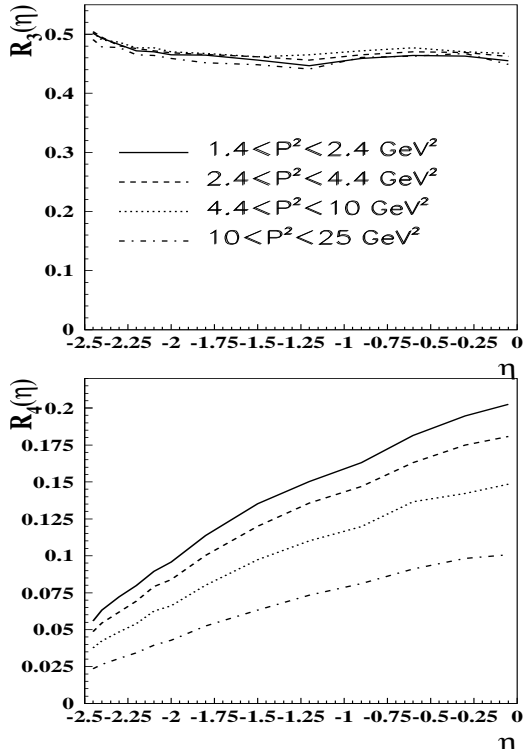

Figure 5. *Nontriviality fractions $R_3$ and $R_4$.*

photon contributions can be characterized [5] by the "nontriviality fractions" $R_3$, $R_4$

$$R_3 \equiv \frac{q^{\text{PL}} \otimes \sigma_q^{\text{res}}(O(\alpha_s^3)) + G^{\text{PL}} \otimes \sigma_G^{\text{res}}(O(\alpha_s^2))}{\sigma^{\text{res}}},$$

$$R_4 \equiv \frac{G^{\text{PL}} \otimes \sigma_G^{\text{res}}(O(\alpha_s^3))}{\sigma^{\text{res}}},$$

plotted as functions of $\eta$ and $P^2$ in Fig. 5. Note that at $\eta = 0$ almost 70% of $\sigma^{\text{res}}$ comes from these origins. This fraction rises even further in the region $\eta > 0$, which, unfortunately is difficult to access experimentally.

So far we have discussed the situation in the window $1.4 \leq P^2 \leq 2.4$ GeV$^2$. As $P^2$ increases the patterns of scale and $R_{\text{sep}}$ dependences change very little. On the other hand, there are some noticable changes:

- The unsubtracted direct photon contribu-

---

tions (DIR$_{\text{uns}}$) represent increasing fractions of the full NLO results.

- The relative contribution of pointlike gluons with respect to pointlike quarks decreases.

- As shown in Fig. 5 the nontriviality factor $R_4$ (which comes entirely from pointlike gluons) decreases, whereas $R_3$, which is dominated by pointlike quarks, is almost independent of $P^2$.

All these features of JETVIP calculations reflect the fundamental fact that as $P^2$ rises towards the factorizations scale $M^2 \approx E_T^2$ the higher order effects incorporated in pointlike parts of photonic PDF vanish and consequently the unsubtracted direct results approach the DIR+RES ones. The crucial point is that for pointlike quarks and gluons this approach is governed by the ratio of $P^2/M^2$. The nontrivial effects included in PDF of the virtual photon thus persist for arbitrarily large $P^2$, provided that we stay in the region where $P^2 \ll M^2$.

**Acknowledgments:** We are grateful to G. Kramer and B. Pötter for discussions on the interactions of virtual photons and to B. Pötter for help with running JETVIP.

---

[5]Disregarding the VDM part of resolved contribution which is tiny in our region of photon virtualities.